\documentclass[aps,prl,twocolumn,groupedaddress]{revtex4}

\usepackage{graphicx}
\begin{document}
\title{Phase diagram of the random Heisenberg antiferromagnetic spin-1 chain}
\author{Andreia \surname{Saguia}}
\affiliation{Centro Brasileiro de Pesquisas F\'{\i}sicas \\
Rua Dr. Xavier Sigaud 150 - Urca,  Rio de Janeiro, 22290-180, RJ - Brazil.}
\author{Beatriz \surname{Boechat}}
\author{Mucio A. \surname{Continentino} }
\email{mucio@if.uff.br}
\affiliation{Departamento de F\'{\i}sica - Universidade Federal Fluminense\\
Av. Litor\^anea s/n,  Niter\'oi, 24210-340, RJ - Brazil}

\date{\today}

\begin{abstract}
We present a new perturbative real space renormalization group
(RG) to study random quantum spin chains and other one-dimensional
disordered quantum systems. The method overcomes problems of the
original approach which fails for quantum random chains with spins
larger than $S=1/2$. Since it works even for weak disorder we are
able to obtain the zero temperature phase diagram of the random
antiferromagnetic Heisenberg spin-1 chain as a function of
disorder. We find a random singlet phase for strong disorder and
as disorder decreases, the system shows a crossover from a
Griffiths  to a disordered Haldane phase.

\end{abstract}
\pacs{75.10.Hk; 64.60.Ak; 64.60.Cn}

\maketitle

The study of the effects of disorder on quantum systems is an
actual and important area of research
\cite{MDH}-\cite{mot}. Intensive work
on the last decades has deepened our understanding of the phase
transitions which occur in pure quantum systems \cite{mucio}. Now
the main effort is concentrated in understanding the role of
randomness in these transitions. This gives rise to new and
interesting phenomena as the existence of Griffiths phases
\cite{fisher,igloi1}. In this connection random quantum spin
chains have been intensive investigated. In the pure case their
behavior is well known \cite{fradkin}. Also for spin$-1/2$ quantum
antiferromagnetic chains a perturbative approach developed by Ma,
Dasgupta and Hu ($MDH$) \cite{MDH} and extended by Fisher
\cite{fisher} allows to obtain results which are essentially exact
for this system. The picture which emerges for these chains is
described by a random singlet phase ($RSP$) where spins are
coupled in pairs over arbitrary distances. In the renormalization
group approach this random singlet phase is governed by an
infinite randomness fixed point \cite{fisher,igloi1,mot}. A
straightforward extension of the $MDH$ method for biquadratic
spin$-1$ chains has shown that in the Heisenberg case the
perturbative RG approach may fail even for the case of strong
disorder \cite{bia}. The reason is that in the elimination process
of strong interactions, in which consists the $MDH$ approach,
interactions stronger than those eliminated are generated. This
failure is better demonstrated when the method is extended to
finite temperatures where it gives rise to non-physical behavior
as negative specific heat and so on \cite{bia4}. Several proposals
have been put forward to extend the $MDH$ method for quantum spin
chains, with $S> 1/2$ \cite{fishernovo}-\cite{jolicoeur},
without undisputed success. The challenge in the case of quantum
integer spin chains is particularly exciting as it deals with the
question of the fate of the Haldane phase \cite{haldane} in the
presence of disorder. The existence of a gap in the excitation
spectrum is not sufficient to guarantee the robustness of pure
chain behavior with respect to the effects of disorder. For gapped
biquadratic chains, any amount of disorder drives the system to a
random singlet phase or infinite randomness fixed point
\cite{bia}. If this is not the case for Haldane chains there may
be a unique property of integer chains which confers them a
special stability with respect to the introduction of disorder. In
fact Hida \cite{hida2} using a density matrix renormalization
group approach has not found any evidence for a $RSP$ for spin-1
chains even in the presence of strong disorder, although this is
still a matter of controversy \cite{hyman,kedar}.

\begin{figure}
\includegraphics[angle=0,scale=0.55]{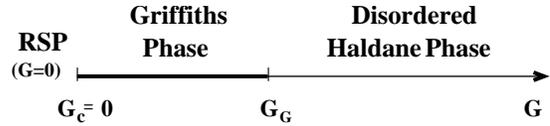}
\caption{ \label{fig1} The phase diagram of the spin$-1$ random
Heisenberg antiferromagnetic chain. Disorder is inversely related
to the gap $G$ of the initial rectangular distribution and $G_G
\approx 0.45$.}
\end{figure}

In this Letter we propose a new approach to the spin$-1$
Heisenberg antiferromagnetic chain which is an improvement of the
traditional $MDH$ perturbative renormalization group method. Our
procedure avoids generation of interactions larger than those
eliminated {\em even for weak disorder}. This provides a unique
opportunity to obtain the phase diagram of the quantum, spin$-1$
Heisenberg chain  as a function of disorder which is shown in
Fig.~\ref{fig1}. We consider here rectangular distributions of
antiferromagnetic interactions
$P(J)=(1/(1-G))\Theta(1-J)\Theta(J-G)$. The gap $G$ is a measure
of the amount of disorder being inversely related to it . We find
for strong disorder, $G=0$, a random singlet phase. As $G$
increases, i.e., disorder decreases, there is a Griffiths phase
characterized by exponents which depend on the distance to the
infinite randomness fixed point at $G=0$. This Griffiths phase
extends up to $G \approx 0.45$ and for weak disorder there is a
disordered Haldane phase.

The $MDH$ method consists in finding the strongest interaction
($\Omega$) between pairs of spins in the chain (see Fig.2a) and
treating the coupling of this pair with their
 neighbors ($J_1$ and $J_2$) as a
perturbation. For a chain of
 spins $S=1$, after elimination of the strongest
coupled pair, the
 new coupling between their neighbors is given by,

\begin{equation}
\label{1} J^{\prime}=\frac{4}{3}\frac{J_1 J_2}{\Omega}
\end{equation}

The factor, $(4/3)>1$, in this equation is the source of the
failure of perturbation theory. Let us assume, for example, that the largest
of
 the neighboring couplings  ($J_1$, $J_2$) to the strongest
interaction $\Omega$ in the chain is $J_1$. If $J_1
> (3/4) \Omega$ than the new effective interaction $J^{\prime}$ is
necessarily larger than one of those eliminated, in this case,
than the weaker one $J_2$.

\begin{figure}
 \includegraphics[scale=0.5]{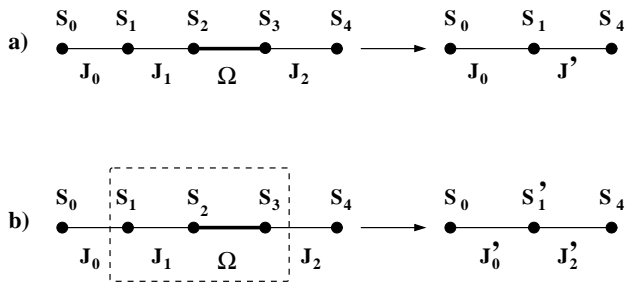}
 \caption{\label{fig2}The two elimination procedures as described in the text ($J_1 > J_2$).}
\end{figure}

Our generalization of the $MDH$ method consists in either of the
following procedures (Fig.1). If the largest neighboring
interaction to $\Omega$, $J_1 < (3/4) \Omega$, then we eliminate
the strongest coupled pair obtaining an effective interaction
between the neighbors to this pair which is given by Eq.~\ref{1}.
This effective interaction is always smaller than those
eliminated.

\begin{figure}
\centering
\includegraphics[angle=0,scale=0.45]{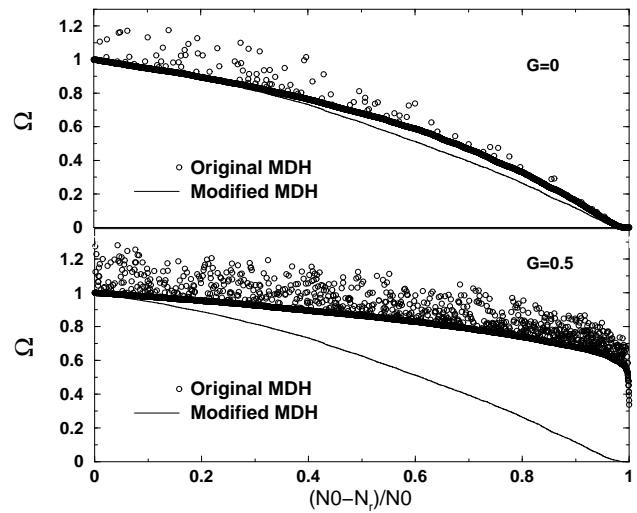}
 \caption{\label{fig3} Evolution of the cut-off of the random chain, as a function of
 the fraction of eliminated spins, under
the two renormalization
 procedures discussed in the text. Note that the naive $MDH$ process generates
interaction larger
 than those eliminated, even for strong disorder ($G=0$). This never occurs in the new procedure used here. }
\end{figure}

Now suppose $J_1 > (3/4) \Omega$ ($J_1 > J_2$). In this case, we
consider the {\em trio} of spins $S=1$ coupled by the two
strongest interactions of the trio, $J_1$ and $\Omega$ and solve
it exactly (see Fig.~2b). The ground state of this trio of spins
$S=1$ is a degenerate triplet and it will be substituted by an
effective spin$-1$ interacting with its neighbors  through new
renormalized interactions obtained by degenerate perturbation
theory. This procedure which implies diagonalizing the $27X27$
matrix of the trio is carried out {\em analytically}. This is
important for obtaining results on large chains and to deal with
the large numbers of initial configurations that we use. These
procedures guarantee that we always comply with the criterion of
validity of perturbation theory and {\em never}, an interaction
larger than those eliminated is generated \cite{note} as shown in
Fig.~\ref{fig3}. Notice from this figure that even for the strong
disorder case, with no gap in the original distribution of
exchange couplings, the simple $MDH$ procedure fails.

\begin{figure}
\centering
 \includegraphics[angle=0,scale=0.35]{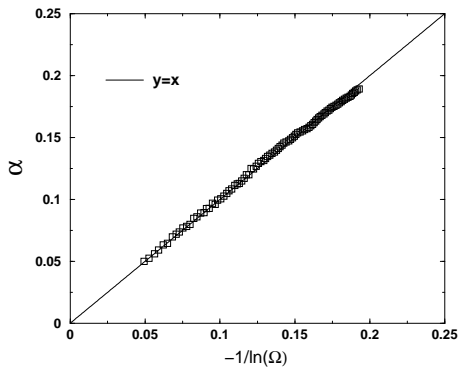}
 \caption{\label{fig4} Exponent \protect$\alpha$ of the fixed point,
power law distribution, Eq.~\ref{2}, as a function of the cut-off
\protect$\Omega$ in the low energy limit.}
\end{figure}

We first consider the strong disorder case $G=0$. This corresponds
to the quantum critical point of the phase diagram where the
system flows to an {\em infinite randomness fixed point}. This
becomes clear when we consider  the fixed point form of the
probability distribution of interactions. This is given by,
\begin{equation}
\label{2}
P(J)=\frac{\alpha}{\Omega}\left(\frac{\Omega}{J}\right)^{1-\alpha}
\end{equation}
The exponent $\alpha$ as a function of the cut-off $\Omega$ is
shown in Fig.~\ref{fig4}. It varies as,
\begin{equation}
\label{3} \alpha=\frac{-1}{\ln \Omega}.
\end{equation}
This behavior characterizes the strong disorder case, $G=0$, as a
random singlet phase \cite{fisher}.

\begin{figure}
\centering
\includegraphics[angle=0,scale=0.35]{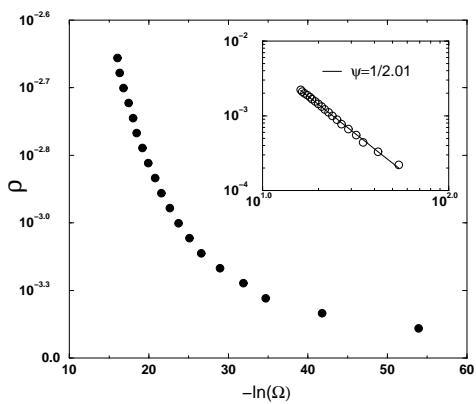}
 \caption{\label{fig5} The density of active spins as a function of the
cut-off in the low energy limit.
  The inset shows the expected behavior for a
random singlet phase with the
 exponent $\psi \approx 1/2$.}
\end{figure}

Further evidence for a random singlet phase at $G=0$ is obtained
considering the fraction of remaining active spins $\rho$
as a function of the energy scale set by
the cut-off $\Omega$ \cite{fisher}. This relation introduces a new exponent
$\psi$ which is defined by,
\begin{equation}
\rho=\frac{1}{L}=\frac{1}{|\ln\Omega|^{1/\psi}}. \label{4}
\end{equation}
It also establishes the connection between the characteristic
length $L$ and the energy scale $\Omega$ for the case of
logarithmic scaling. This is an extension of the usual definition
of a dynamic exponent ($\Omega^{-1} \propto \tau \propto L^{z}$).
In Fig.~\ref{fig5} we show the density $\rho=1/L$ as a function of
the cut-off. From this expression we extract the exponent $\psi$
which takes the value $\psi=1/2$ characteristic of the random
singlet phase \cite{fisher}.

\begin{figure}
\centering
{\includegraphics[angle=0,scale=0.23]{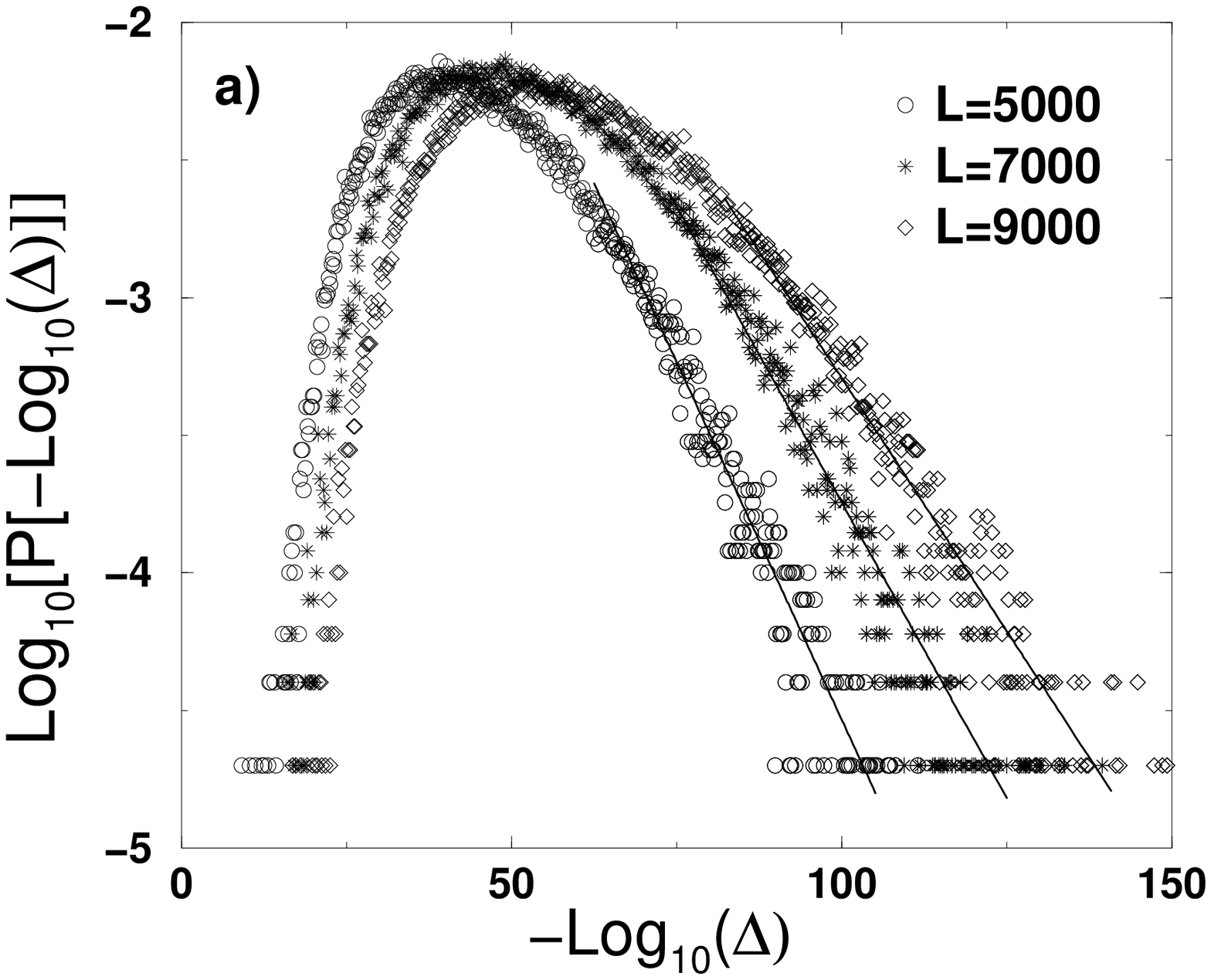}}
{\includegraphics[angle=0,scale=0.23]{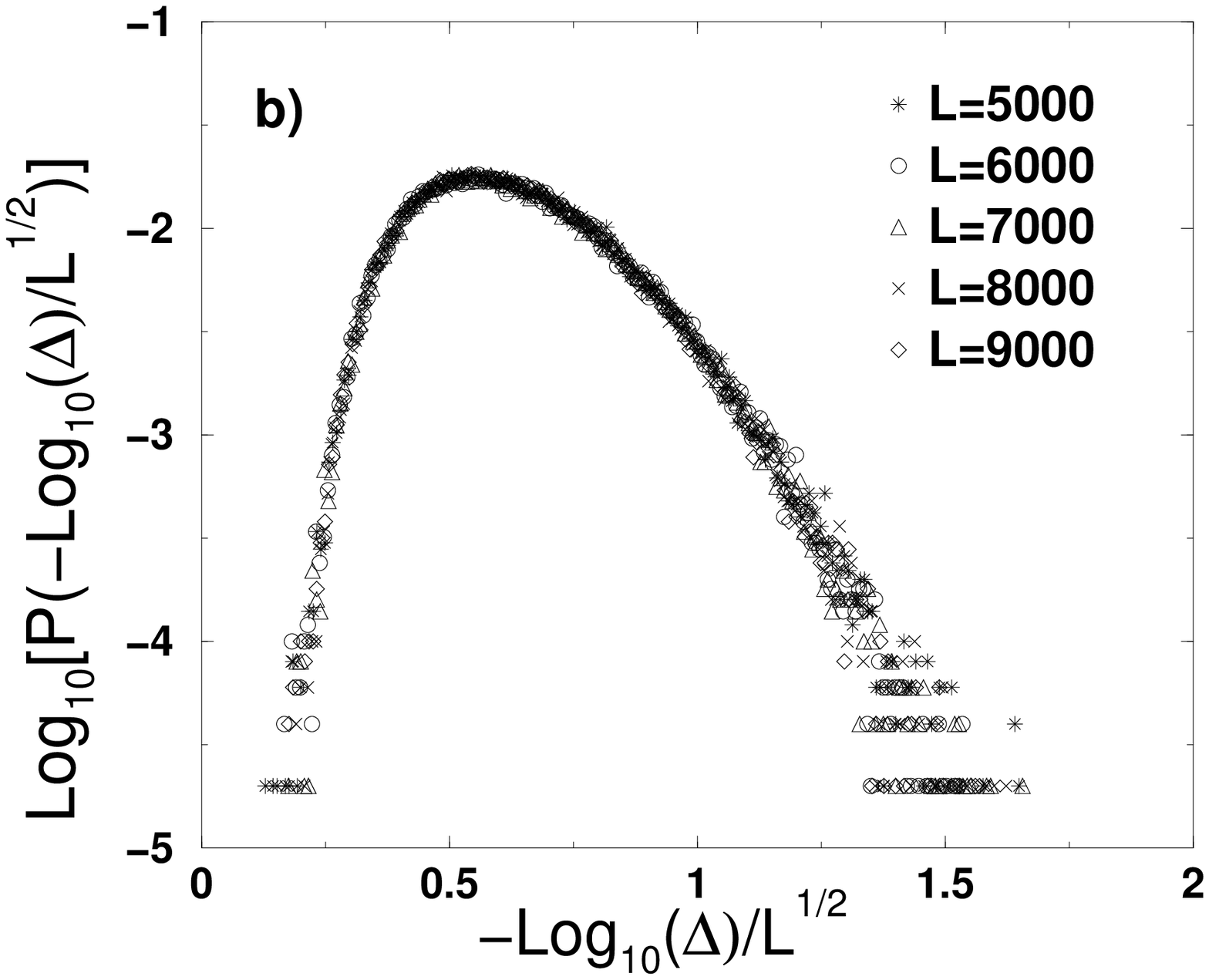}}
\caption{ \label{fig6}  a) Probability distribution of the first gap at
the transition point, $G=0$. The distribution become broader and broader
with L, which signals infinite randomness behavior. b) Scaling plot for
the gap distributions.
The collapse of the curves is obtained for $\psi = 1/2$ as expected
for a random singlet phase.}
\end{figure}

Finally we calculate  the distribution of first gaps at $G=0$
\cite{igloi1, igloi2}. This is obtained starting from a given
configuration of random interactions for a chain of size $L$ and
eliminating the spins, as described above, until a single pair
remains. The interaction between these remaining spins yields the
first gap $\Delta$ for excitation. Implementing this procedure for
a large number of initial random configurations for chains of
different sizes $L$ yields the distributions $P_L(\log \Delta)$
shown in Fig.~\ref{fig6}. We considered over $10^4$ initial
configurations to obtain the gap distributions. The widths of
these distributions increase without limit as the sizes $L$ of the
chains increase, as expected for an infinite randomness fixed
point

According to the scaling form relating energy and length,
Eq.~\ref{4}, we expect that the distribution $P(-\log \Delta /
L^{\psi})$ will present a universal behavior, independent of the
size $L$ of the chains when plotted versus the variable $-\log
\Delta / L^{\psi}$. This is indeed the case for $G=0$ as shown in
Fig.~\ref{fig6} for the $RSP$ exponent $\psi=1/2$.

\begin{figure}
\centering
{\includegraphics[angle=0,scale=0.45]{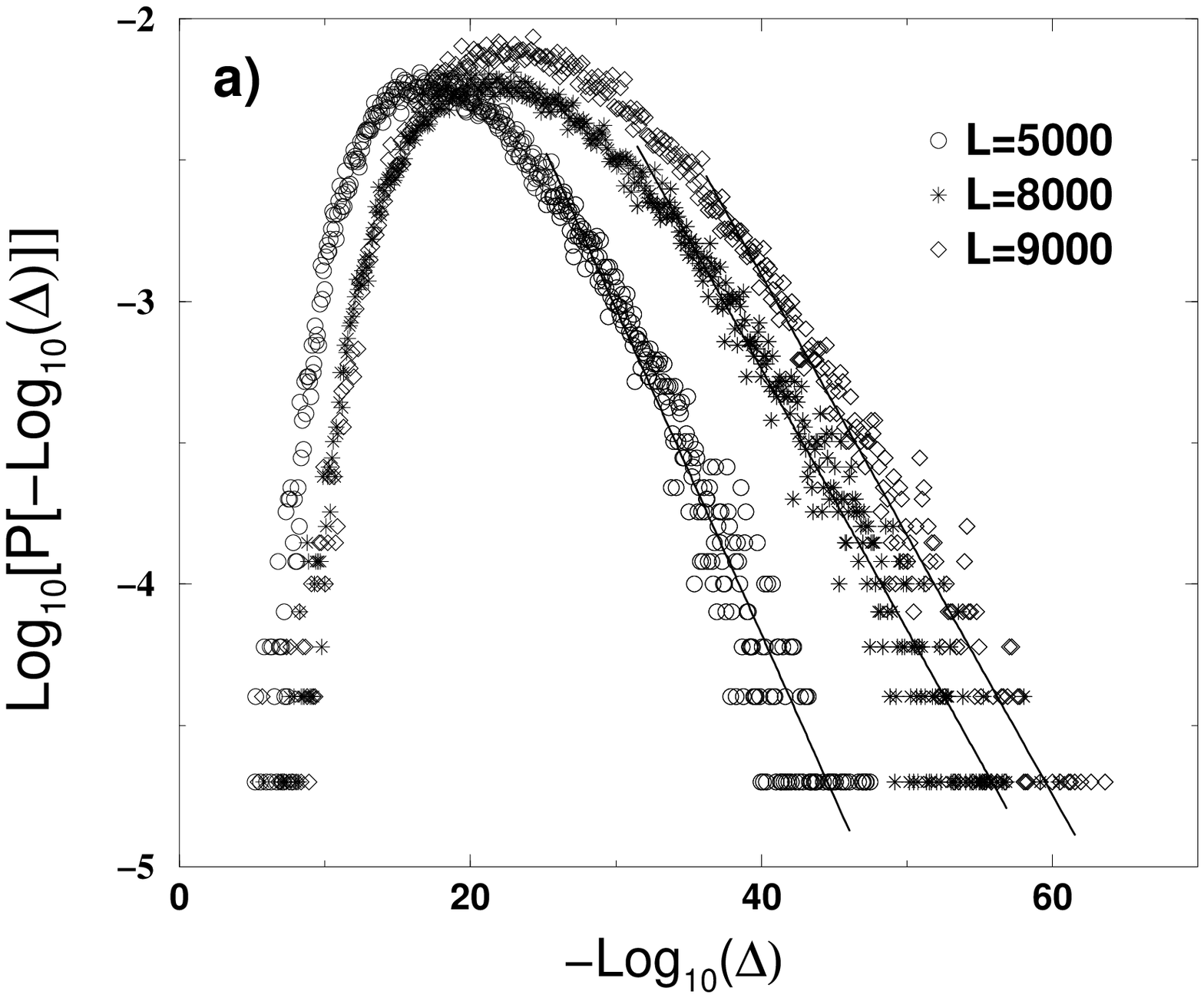}}
{\includegraphics[angle=0,scale=0.45]{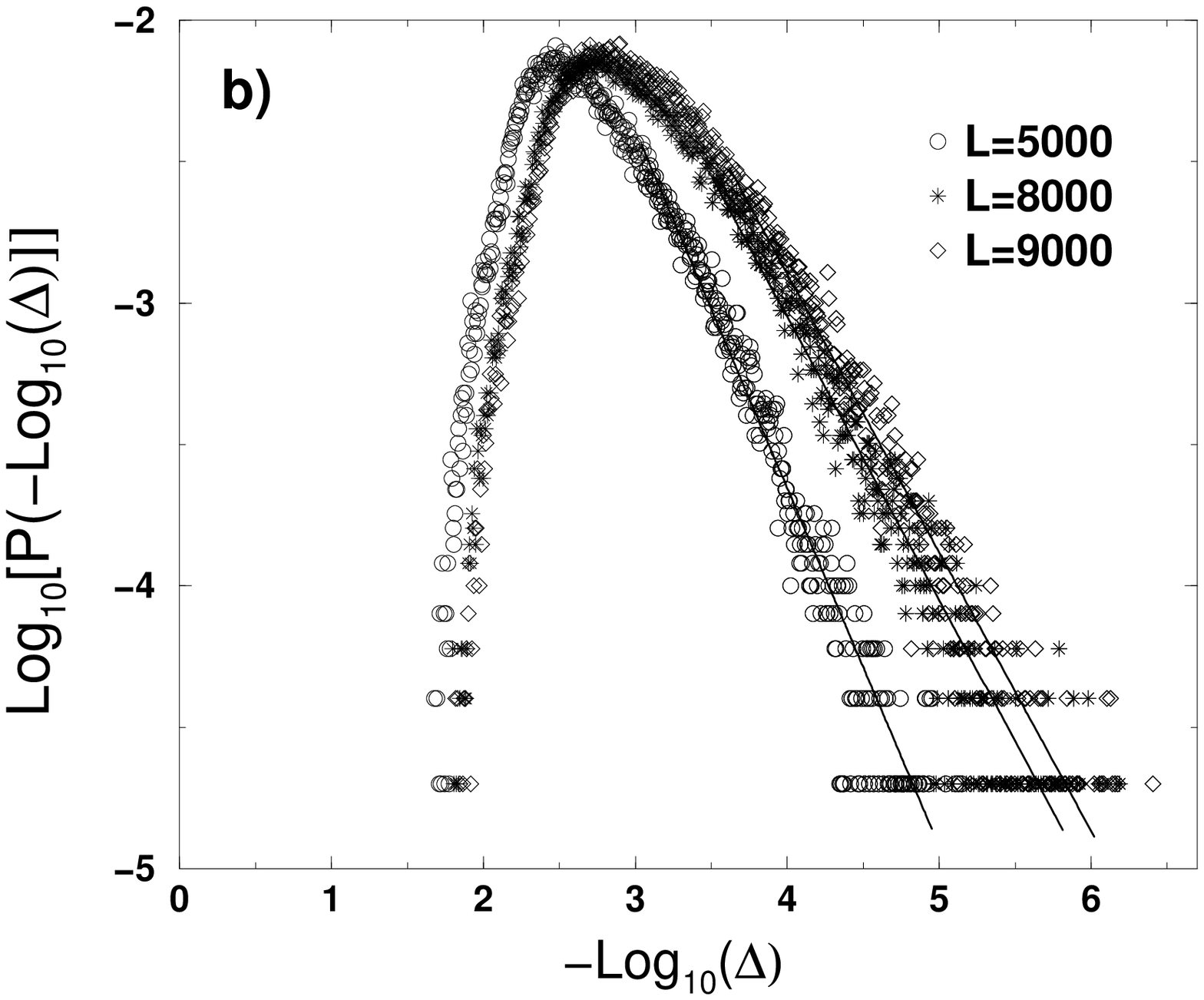}}
{\includegraphics[angle=0,scale=0.45]{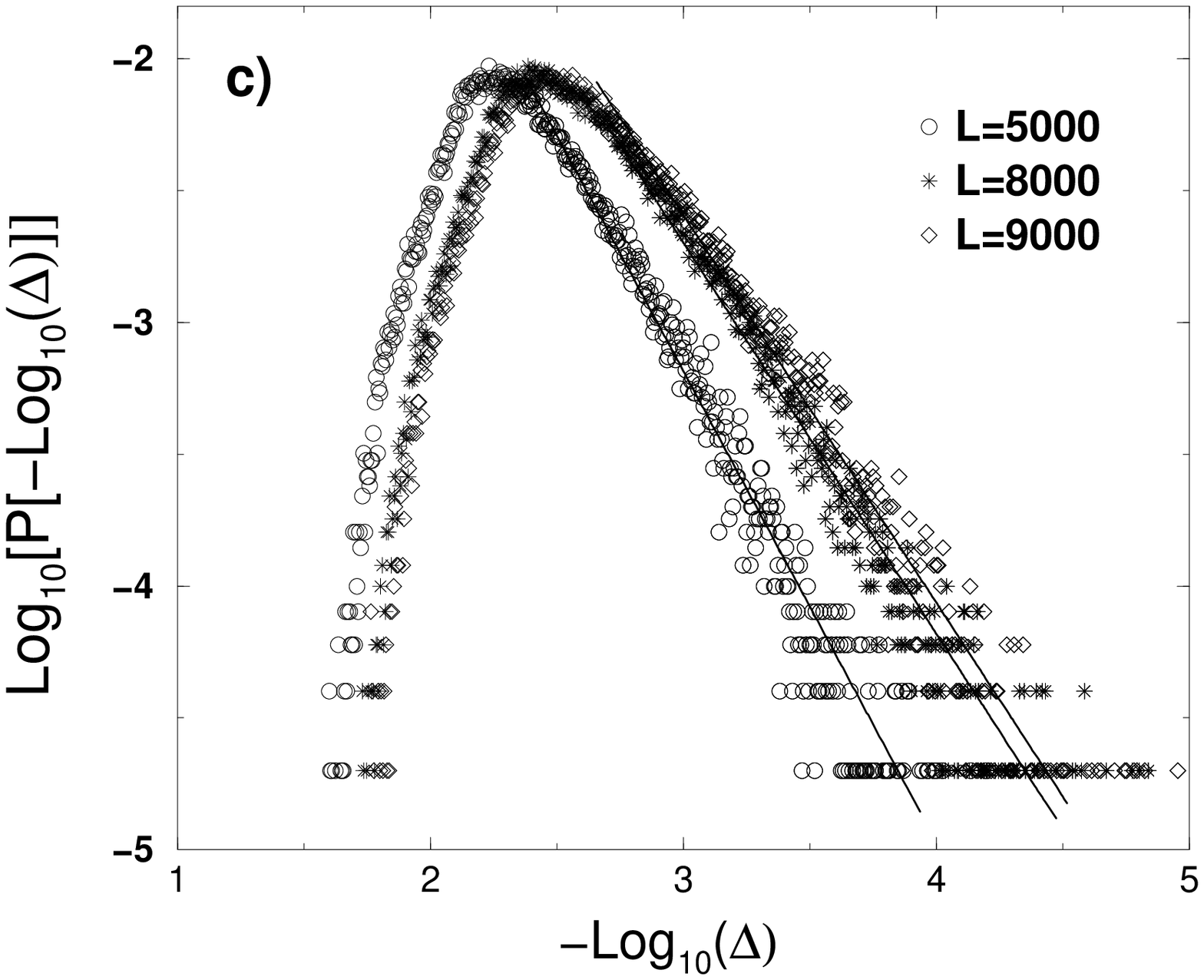}}
\caption{ \label{fig7} Probability distributions of the first
gap obtained from initial rectangular distributions of couplings with a gap
$G$ and different systems sizes $L$. For clarity not all values of $L$ are
shown. The solid lines represent best fits to the form $\log_{10}[P(-\log_{10}
\Delta)]=A_L-\frac{1}{Z_L} \log_{10} \Delta$. a) $G=0.1$, $Z_{5000}=8.69$,
$Z_{6000}=8.70$, $Z_{7000}=10.45$, $Z_{8000}=10.85$ and $Z_{9000}=10.87$. b)
$G=0.45$, $Z_{5000}=0.79$, $Z_{6000}=0.81$, $Z_{7000}=0.96$, $Z_{8000}=1.0$
and $Z_{9000}=1.01$. c) $G=0.5$, $Z_{5000}=0.55$, $Z_{6000}=0.58$,
$Z_{7000}=0.63$, $Z_{8000}=0.67$ and $Z_{9000}=0.68$.}

\end{figure}

We now decrease disorder increasing the gap in the exchange
distribution. In Figure~\ref{fig7} we show the first gap
distributions for different degrees of disorder as characterized
by the gaps $G$ in the initial distribution of interactions. In
all cases that we have investigated with $G \ne 0$, we find that
the first gap distributions saturate at low energies in a form
described by the expression, $P(\log \Delta) \sim \Delta^{1/Z}$
for $\Delta \rightarrow 0$. The {\em dynamic exponent} $Z$ becomes
independent of $L$ for $L$ sufficiently large. We have to consider
large chains in order to observe this effect. We find $Z_{\infty}
\sim 10.87$, $Z_{\infty} \sim 1.01$ and $Z_{\infty} \sim 0.68$ for
$G=0.1$, $G=0.45$ and $G=0.5$, respectively. From these values of
the dynamic exponent we can deduce the existence of a Griffiths
phase extending up to $G_G \approx 0.45$ where the dynamic
exponent reaches the value $Z=1$. For values of the gap $G>G_G$,
i.e., small disorder, the dynamic exponent $Z  < 1$. The
distribution of first gap for excitations, from which low
temperature thermodynamic properties can be deduced, implies that
$Z>1$ is required to obtain a singular behavior for these
quantities with decreasing temperature. Consequently at $G_G$
there is a significant change in the nature of the thermodynamic
behavior of the system.  The phase for $G
>G_G$ is a disordered Haldane phase with a pseudo-gap in the excitation spectrum.
The thermodynamic behavior along the phase diagram will be
explored in a future publication \cite{saguia}.

We have generalized the $MDH$ perturbative renormalization group.
This method was known to fail in the case of random quantum chains
with spins $S>1/2$ as it generates couplings which are larger than
those eliminated, signalling the breakdown of perturbation theory.
Taking into account larger clusters and treating them exactly we
were able to circumvent this problem. For the important case of
the spin$-1$ random Heisenberg antiferromagnetic chain our
elimination procedure gives rise to interactions which are always
smaller than those eliminated, even for weak disorder. This allows
us to obtain the phase diagram of this system as a function of
disorder. We find for initial rectangular distributions with no
gap, $G=0$, a random singlet phase similar to that found in
spin$-1/2$ chains. For finite values of the gap $G$, the first gap
distribution $P_L(\log \Delta)$ becomes, in the low energy limit,
independent of $L$ for sufficiently large chains and is
characterized by a dynamic exponent $Z$ which depends on how far
the system is from the infinite randomness fixed point at $G=0$.
This Griffiths phase is associated with a dynamic exponent $Z \ge
1$ and is limited by the fixed point at $G=0$ (where $Z= \infty$)
and by $G_G \approx 0.45$, where $Z$ attains the value $Z=1$. For
larger values of the gap, i.e., small disorder the system presents
a disordered Haldane phase with a pseudogap in the spectrum for
excitations.

\begin{acknowledgments}

We would like to
thank Conselho Nacional de De\-sen\-vol\-vi\-men\-to Cient{\'{\i}}fico
  e
Tecnol\'ogico-CNPq-Brasil (PRONEX98/MCT-CNPq-0364.00/00), Fundac\~ao de Amparo
a Pesquisa do Estado do Rio de Janeiro-FAPERJ for partial financial
support.

\end{acknowledgments}


\begin{thebibliography}{99}
\bibitem{note} This is checked at every elimination.
\end{thebibliography}
\end{document}